\documentstyle[12pt,epsf]{article}
\newcommand{\la}[1]{\label{#1}}

\newlength{\numlen}

\newcommand{\be}{\begin{equation}}
\newcommand{\ee}{\end{equation}}
\newcommand{\ba}{\begin{eqnarray}}
\newcommand{\ea}{\end{eqnarray}}
\newcommand{\rmi}[1]{{\mbox{\scriptsize #1}}}

\newcommand{\bfx}{\mbox{\bf x}}
\newcommand{\bfb}{\mbox{\bf b}}

\newcommand{\roots}{$\sqrt{s}$}
\newcommand{\fm}{\rm fm}
\newcommand{\gev}{\rm GeV}

\def\lsim{\raise0.3ex\hbox{$<$\kern-0.75em\raise-1.1ex\hbox{$\sim$}}}
\def\gsim{\raise0.3ex\hbox{$>$\kern-0.75em\raise-1.1ex\hbox{$\sim$}}}

\begin{document}
 
\begin{titlepage}
\begin{flushright}
CERN-TH/96-259\\
October 10, 1996\\
\end{flushright}
\begin{centering}
\vfill

{\bf 
          BARYON-TO-ENTROPY RATIO IN VERY HIGH ENERGY
                     NUCLEAR COLLISIONS 
}

\vspace{0.5cm}
 K.J.Eskola$^{\rm a,b}$\footnote{kari.eskola@cern.ch} and
 K. Kajantie$^{\rm a,b}$\footnote{keijo.kajantie@cern.ch}

\vspace{1cm}
{\em $^{\rm a}$ CERN/TH, CH-1211 Geneve 23, Switzerland\\}
\vspace{0.3cm}
{\em $^{\rm b}$ Department of Physics,
P.O.Box 9, 00014 University of Helsinki, Finland\\}

\vspace{1cm}
{\bf Abstract}

\end{centering}

\vspace{0.3cm}\noindent
We compute as a function of rapidity $y$ 
the baryon number carried by quarks and antiquarks
with $p_{\rm T}>p_0\approx$ 2 GeV produced in Pb+Pb collisions at
TeV energies. The computation is carried out in lowest order 
QCD perturbation theory using structure functions 
compatible with 
HERA results. At $p_0=2$ GeV the initial gluon density is
both transversally saturated and thermalised in the sense
that the energy/gluon equals to that of an ideal gas with the 
same energy density. Even at these high energies the initial net 
baryon number density at $y=0$ at $\tau=0.1$ fm will be more 
than the normal nuclear matter density but the baryon-to-entropy 
ratio is only $(B-\bar B)/S\sim 1/5000$. Further evolution of the 
system is discussed and the final baryon-to-entropy ratio is 
estimated. 

\vfill 
\noindent

\end{titlepage}
\section{Introduction} 

The formation of QCD plasma in central collisions  of large nuclei 
in the TeV energy range can be described as the formation of a 
large number of partons (gluons and quarks) with transverse 
momentum $p_{\rm T}>p_0$, where $p_0$ parametrises the lower limit 
of the $p_{\rm T}$ of the partons included in the computation 
\cite{bm}. The number of these partons in an average collision can 
be estimated in QCD perturbation  theory \cite{ekl},\cite{ekr}.
It scales $\sim A^{4/3}$ and for $p_0\approx$ 2 GeV its value is 
easily ${\cal O}(1000)$ per unit of rapidity in the central rapidity 
region. This process takes place in a time $1/p_0\approx$ 0.1 fm/c 
and  it is thus equal to the initial formation of QCD plasma. Making 
predictions for the final observables requires modelling the further 
evolution of the system either hydrodynamically \cite{eg,ms,grz} or in 
various kinetic models  \cite{geiger92}-\cite{venus}.

For a thermal system the computation of the overall multiplicity 
of the initial partons gives the entropy, associated with
the temperature $T(t,\bfx)$, and a computation of the
quark and antiquark number gives the net quark number,
associated with the chemical potential $\mu(t,\bfx)$.
The purpose of this article is to compute in more detail the 
initial amounts of gluons, quarks and antiquarks with 
$p_{\rm T}>p_0$ and from this to estimate the final 
baryon-to-entropy ratio.

At LHC energies, \roots = 5500 GeV, the gluons are by far 
the dominant component of the initial state. In fact, for
$p_0=2$ GeV we shall see that QCD perturbation theory implies 
that the gluonic subsystem has two remarkable properties: it
is both thermalised, in the sense that  its energy/gluon is 
the same as in a thermally equilibrated ideal gas with the 
corresponding energy density, and transversally saturated 
\cite{glr,bm,wg91} in the sense that the total transverse area 
occupied by the gluons equals the total nuclear area. Thus, 
although $p_0$ in principle is  a parameter constrained by  
soft physics \cite{wang91,sz}, its value 2 GeV gets a well defined 
dynamical significance in  heavy ion collisions at very high energies.

After the initial production the system goes through various
expansion and phase transition stages. A natural first assumption
is to assume that the expansion is adiabatic, so that both
the total entropy $S$ and the net baryon number $B-\bar B$
in a unit of rapidity are constant. This, in particular, implies 
that $E_T$ in a unit of rapidity decreases rapidly, the work done
against expansion transfers $E_T$ to larger rapidities from the central
unit. We shall analyse this decrease, as well as possible increase
of $S$ and $B-\bar B$ due to dissipation and non-perturbative source terms 
in some detail. 

Concerning $B-\bar B$, the system produced in pp collisions is 
different from the one in $AA$  collisions. In pp the color 
correlations (strings) between individual quarks (antiquarks)  
produced in the central rapidity  and the (di-)quarks left in the 
fragmentation regions are essential  for baryon production 
\cite{fritiof,venus,wg91,hijing}. The system  is dilute and 
essentially free streaming, and the (net) baryon number 
remains mostly in the fragmentation regions,  while the mid-rapidity 
quark(antiquark) usually becomes an  ingredient of a meson. 
Therefore, in pp collisions the initially  produced $q-\bar q$ 
distribution in the mid-rapidity range does not  directly reflect the 
final net baryon distribution. In heavy ion collisions, however, 
the situation  should be different.  At very high energies, 
the produced parton system in the central rapidity region is 
extremely dense, and its further evolution depends rather on the 
secondary collisions of partons (and strings  as well) than on 
the direct color correlations with the fragmentation 
regions.  This dynamics is built in hydrodynamical expansion
equations, which imply in the adiabatic case that $B-\bar B$ in
a unit of rapidity is constant.

At \roots = 5500 GeV the total number of quarks and antiquarks
will be seen to be initially only 10\% of that of gluons. 
The net quark number is much less but still calculable. 
In fact, we estimate that the  initial $(B-\bar B)/S\approx1/5000$  
at $\tau=0.1$ fm which then may increase by up to a factor 4, 
while  in the \roots=20 GeV  range numbers of the order of 
1/20...1/50 have been observed  \cite{lthsr,crss}. Our result 
is in qualitative agreement with  the expectation 
that at very high  energies the baryon number of the initial 
nuclei goes to large  forward and backward rapidities leaving 
the central region  essentially depleted of net baryon number 
\cite{hs,uh}.  Experimental data at AGS ($\sqrt{s}$=5 GeV)
and SPS ($\sqrt{s}$ = 17 GeV) \cite{leaddata}  have shown that 
nuclei are much more efficient in stopping baryons than pp 
collisions and it has been recently argued in a gluon field 
model of baryons \cite{kharzeev} that this would persist to
LHC energies ($\sqrt{s}$ in the TeV range). Also effects on 
the baryon-to-entropy ratio due  to event-by-event fluctuations 
of the net baryon number have been discussed recently \cite{spieles}.

This approach of computing gluon and quark production stretches
QCD perturbation to its limits and the sources of error are
obvious: next-to-leading order (NLO) corrections with the present 
acceptance criteria have not been computed, there are uncertainties
associated with small-$x$ gluon shadowing and with modeling nuclear 
effects for gluons in general,  the  quantitative results will
somewhat depend on the parameter $p_0$ defining the lower limit of 
the $p_{\rm T}$ of quarks and gluons included.  We offer the results 
for what they are worth; they will anyway set  the stage for further 
refinements.

\section{The formulas}

The lowest order (LO) formula for inclusive production of partons in 
pp collisions can be found  {\it e.g.} in Refs. \cite{XS}. 
Our main focus is on the production of semihard partons, defined 
by $p_{\rm T} > p_0$. The cross section for each flavour $f$ 
can be written as follows:

\begin{eqnarray}
\nonumber
\frac{d\sigma^f}{dy}(\sqrt s,p_0) = 
\int dp_{\rm T}^2dy_2\, \sum_{ij\atop {\langle kl\rangle}}  
x_1f_{i/p}(x_1,Q^2)\, x_2 f_{j/p}(x_2,Q^2)\times \\
\times \biggl[
\delta_{fk}\frac{d\hat\sigma}{d\hat t}^{ij\rightarrow kl}
\hspace{-0.7cm}(\hat t,\hat u)
+
\delta_{fl}\frac{d\hat\sigma}{d\hat t}^{ij\rightarrow kl}
\hspace{-0.7cm}(\hat u,\hat t)
\biggr]\,
\frac{1}{1+\delta_{kl}},\la{rate}
\end{eqnarray}
where the fractional momenta of incoming partons are
\begin{equation}
x_1 = \frac{p_{\rm T}}{\sqrt s}({\rm e}^{y}+{\rm e}^{y_2})\,\,\,\,\,
{\rm and}\,\,\,\, 
x_2 = \frac{p_{\rm T}}{\sqrt s}({\rm e}^{-y}+{\rm e}^{-y_2}),
\end{equation}
the region of integration is 
\begin{equation}
p_0^2\le p_{\rm T}^2\le \frac{s}{4\cosh^2y}
\,\,\,\,\, {\rm and}\,\,\,\,\, 
\ln(\frac{\sqrt s}{p_{\rm T}}-{\rm e}^{-y})\le y_2 \le 
\ln(\frac{\sqrt s}{p_{\rm T}}-{\rm e}^{y}),
\end{equation}
where $y$ is fixed so that
\be
|y|\le\ln\biggl(\frac{\sqrt s}{2p_0}+\sqrt{\frac{s}{4p_0^2}-1}\,\biggr)
\ee
and the invariants at the parton level are 
\begin{equation}
\hat t = -p_{\rm T}^2[1+{\rm e}^{-(y-y_2)}]\,\,\,\,\,
{\rm and}\,\,\,\, 
\hat u = -p_{\rm T}^2[1+{\rm e}^{(y-y_2)}].
\end{equation}
The sum over initial states includes all combinations of two 
parton species: 
\begin{equation}
ij = gg, gq, qg, g\bar q, \bar qg, qq, q\bar q,\bar qq, \bar q\bar q,
\end{equation}
while the final states consist of all pairs (without a 
mutual exchange):
\begin{equation}
\langle kl\rangle =  gg, gq, g\bar q, qq, q\bar q, \bar q\bar q.
\end{equation}
The factor $1/(1+\delta_{kl})$ is a statistical factor for
identical particles in the final state. 
In the initial state, three (anti)quark flavours are taken 
into accout: $q = u,d,s$, 
and, in the final state four (massless) flavours: $q = u,d,s,c$.
In the running $\alpha_s(p_{\rm T})$ we, accordingly, use $N_f=4$. Since
\begin{equation}
\frac{d\hat\sigma}{d\hat t}^{ij\rightarrow kl}
\hspace{-0.7cm}(\hat t,\hat u)
=
\frac{d\hat\sigma}{d\hat t}^{ji\rightarrow kl}
\hspace{-0.7cm}(\hat u,\hat t)
\end{equation} 
there are only eight different types of subprocess cross
sections:
\begin{eqnarray}
\nonumber  gg         \rightarrow gg      &&  gq       
\rightarrow gq\\
\nonumber  gg         \rightarrow q\bar q &&  q_iq_j   
\rightarrow q_iq_j, \,\, i\neq j\\
\nonumber  q_iq_i     \rightarrow q_iq_i  &&  q_i\bar q_i
\rightarrow q_j\bar q_j, \,\, i\neq j\\ 
\nonumber  q_i\bar q_i\rightarrow q_i\bar q_i   &&  q_i\bar q_i
\rightarrow gg,
\end{eqnarray}
which can be found, e.g., in refs. \cite{XS}.
 
For the parton densities $f_{j/p}$, we use the set GRV 94 LO 
\cite{GRV94} with a scale 
choice $Q=p_{\rm T}$. With this set we can consistently extend 
our lowest order computation even below $p_{\rm T}=2$ GeV.

Summation over final state parton flavours $g,u,d,s,c$ now gives
the production cross sections 
$d\sigma^g/dy,d\sigma^q/dy,d\sigma^{\bar q}/dy$ of gluons, quarks
and antiquarks with $p_{\rm T}>p_0$ in a pp collision. These are
normalised so that
\be
\int dy \frac{d\sigma^f}{dy} \equiv 2\sigma_{\rm hard}^f(\sqrt s,p_0)
\ee
and
\be
\sigma_{\rm hard}^g(\sqrt s,p_0)+
\sigma_{\rm hard}^q(\sqrt s,p_0)+
\sigma_{\rm hard}^{\bar q}(\sqrt s,p_0) \equiv 
\sigma_{\rm hard}(\sqrt s,p_0),
\ee
where $\sigma_\rmi{hard}(\sqrt{s},p_0)$ is the cross section for one
(semi)hard collision and
the factor 2 comes from two partons in one (semi)hard event.
Since we will be especially interested in the central rapidity region,
$|y|\le0.5$, let us also define for later use 
\be
\sigma_{\rm hard}^f(\sqrt s,p_0,|y|\le0.5) \equiv \frac{1}{2}
\int_{-0.5}^{+0.5} dy \frac{d\sigma^f}{dy}
\ee

The next-to-leading (NLO) order corrections to eq.(\ref{rate}) can
be large and UA1 minijet data indicates \cite{ekl} that
a K factor of about 2.5 is needed at those energies. NLO
jet analysis shows \cite{ew/hp} that at higher energies the factor is
somewhat less, K$\approx1.5$. To
get a conservative estimate we shall use K = 1.

After the cross sections are computed on the pp level, they
are transformed to the number $\bar N$ of events of the
type described by $\sigma$ in an average central (zero impact
parameter) A+A collision by multiplying by the nuclear
overlap function $T_\rmi{AA}(\bfb)$ at $\bfb=0$: $
\bar N=\sigma T_\rmi{AA}(\bfb=0)$. For example, the number
distributions of produced flavour $f$ are obtained as
\be
{dN^f\over dy}=T_{AA}(\bfb=0){d\sigma^f\over dy}, \label{number}
\ee
and the total number of partons of flavor $f$ within $|y|\le 0.5$ as
\be
\bar N_{AA}^f({\bf b=0},\sqrt s,p_0) = 
2\,T_{AA}({\bf 0})\sigma_{\rm hard}(\sqrt s,p_0,|y|\le0.5)
\ee
Numerically $T_\rmi{AA}(0)\approx A^2/\pi R_A^2$ and we shall
use the value
\be
T_\rmi{PbPb}(0)={32\over{\rm mb}}. \label{overlap}
\ee
This procedure is equivalent to neglecting nuclear shadowing, {\it i.e.}
assuming that $f_{i/A}(x)=Af_{i/p}(x)$. When estimating
various quantities we shall further use
\be
R_\rmi{Pb}=6.54\,\,{\rm fm},\quad \pi R_\rmi{Pb}^2=134\,\,{\rm fm}^2,
\quad V_i=\pi R_\rmi{Pb}^2\Delta y/p_0=13.4\,\,{\rm fm}^3,
\label{radii}
\ee
where $V_i$ is the initial volume at the time $\tau_i=1/p_0$
(with $p_0=2$ GeV) when the quarks and gluons were formed. 
Note that for the longitudinal width for the volume near 
the central rapidity,  we have approximated  
$\Delta z\approx \tau_i\Delta y$, {\it i.e.}
we have neglected effects of the longitudinal spatial spread 
of the beam-partons.

A lower limit for the parameter $p_0$ is obtained from the
transverse saturation criterion \cite{bm,glr,wg91}. The partonic
subsystem is saturated if 
the total transverse area occupied by the produced partons
= $\bar N_{AA}(\bfb=0)\pi/p_0^2$ is larger than the total nuclear area
= $\pi R_A^2$. For Pb+Pb and 
$\bar N_{\rm PbPb}(\bfb=0)=2\sigma_{\rm hard}(\sqrt s,p_0,|y|\le0.5) 
T_\rmi{PbPb}({\bf 0})$ 
this converts to
\be
{\sigma_{\rm hard}(\sqrt s,p_0,|y|\le0.5)\over{\rm mb}}
>67\biggl({p_0\over 2\,{\rm GeV}}
\biggr)^2.
\la{saturation}
\ee

The formulas required for the computation of $E_{\rm T}$-distribution in a 
hard collision with an acceptance region can be found in \cite{ekl}.
We define our acceptance through a ``measurement function'' 
$\epsilon(y)$, which will in the following be chosen as a step function
\begin{equation}
\epsilon(y) = \left\{ \begin{array}{ll}
                      1, & \mbox{if $|y|\le0.5$} \\
                      0, & \mbox{otherwise.}
                      \end{array}
              \right.
\end{equation}
In LO the (mini)jets are produced back-to-back in ${\bf p_{\rm T}}$, and 
the $E_{\rm T}$-distribution in the chosen acceptance
window can now be defined for each flavor $f$ as
\begin{eqnarray}
\frac{d\sigma^f}{dE_{\rm T}} = 
\frac{1}{2}\int dp_{\rm T}^2dy_1dy_2\, \sum_{ij\atop {\langle kl\rangle}}  
x_1f_{i/p}(x_1,Q^2)\, x_2 f_{j/p}(x_2,Q^2)\frac{1}{1+\delta_{kl}}
\times\hspace{3cm}\nonumber \\
\times\biggl\{
\frac{d\hat\sigma}{d\hat t}^{ij\rightarrow kl} \hspace{-0.7cm}(\hat t,\hat u)
\,\delta(E_{\rm T}-
[\delta_{fk}\epsilon(y_1) + \delta_{fl}\epsilon(y_2)]p_{\rm T})
+\nonumber\\
+ \frac{d\hat\sigma}{d\hat t}^{ij\rightarrow kl}\hspace{-0.7cm}(\hat u,\hat t)
\,\delta(E_{\rm T}-
[\delta_{fl}\epsilon(y_1) + \delta_{fk}\epsilon(y_2)]p_{\rm T})
\biggr\},
\end{eqnarray}
with the normalisation
\begin{equation}
\sum_f \int dE_{\rm T}\frac{d\sigma^f}{dE_{\rm T}} = 
\sigma_{\rm hard}(\sqrt s,p_0). 
\end{equation}
As explained in \cite{ekl}, this formulation takes into account that 
of the two minijets produced in the hard collision, there may be one or 
two or none (of flavor $f$) within the acceptance window.  
It is straightforward to obtain the first $E_{\rm T}$-moment for 
each flavor $f$  in the hard collision, 
\begin{eqnarray}
\sigma_{\rm hard}(\sqrt s,p_0)\langle E_{\rm T}^f\rangle = 
\int dE_{\rm T}\frac{d\sigma^f}{dE_{\rm T}} \hspace{7.5cm}\nonumber\\
\hspace{3cm}= \int dp_{\rm T}^2dy_1dy_2\, \sum_{ij\atop {\langle kl\rangle}}  
x_1f_{i/p}(x_1,Q^2)\, x_2 f_{j/p}(x_2,Q^2)\frac{1}{1+\delta_{kl}}
\times \nonumber\\
\times \biggl[
\delta_{fk}\frac{d\hat\sigma}{d\hat t}^{ij\rightarrow kl}
\hspace{-0.7cm}(\hat t,\hat u)
+
\delta_{fl}\frac{d\hat\sigma}{d\hat t}^{ij\rightarrow kl}
\hspace{-0.7cm}(\hat u,\hat t)
\biggr] p_{\rm T}\epsilon(y_1)
\la{sET}
\end{eqnarray}
{\it i.e.} basically as the integral of Eq. (\ref{rate}) over $y$ in 
the accepted $y$-region, weighted by $p_{\rm T}$. 

In an $AA$ collision with an impact parameter ${\bf b}$,
the average transverse energy carried by semihard partons of flavor $f$ 
to the acceptance region becomes then
\be
\bar E_{\rm T}^f({\bf b},\sqrt s,p_0) = 
T_{AA}({\bf b}) \sigma_{\rm hard}(\sqrt s,p_0)\langle E_{\rm T}^f\rangle,
\ee
where $T_{AA}({\bf b})\sigma_{\rm hard}(\sqrt s,p_0)$ is the 
average number of semihard collisions (all $y$) and 
$\langle E_{\rm T}^f\rangle$ is the average transverse energy 
of the flavor $f$ at $|y|\le 0.5$ in each semihard collision.

\begin{figure}[tb]
\vspace*{2cm}
\centerline{\hspace{-3.3mm}
\epsfxsize=10cm\epsfbox{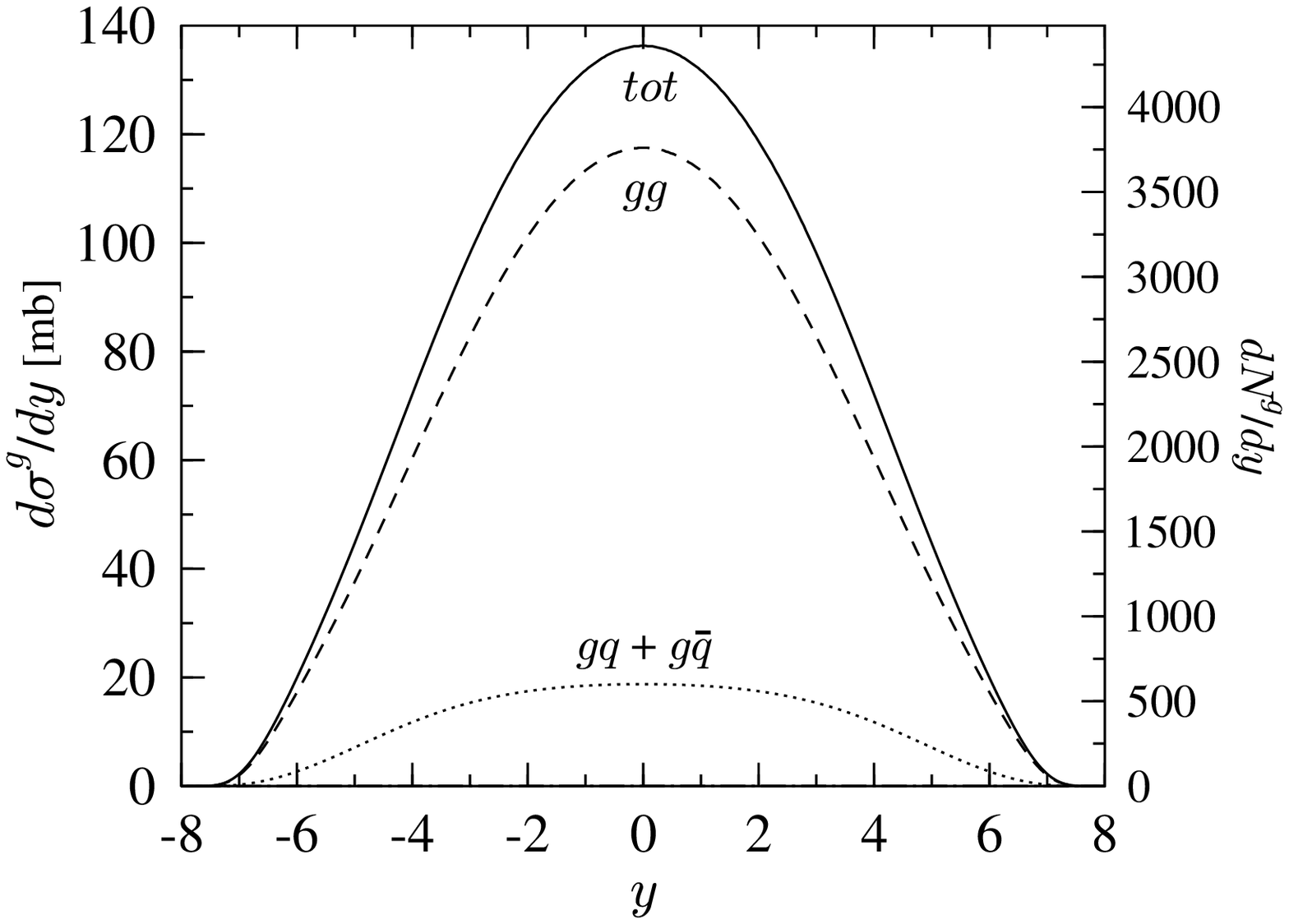}
}
\vspace{-2.4cm}
\centerline{\hspace{-3.3mm}
\epsfxsize=10cm\epsfbox{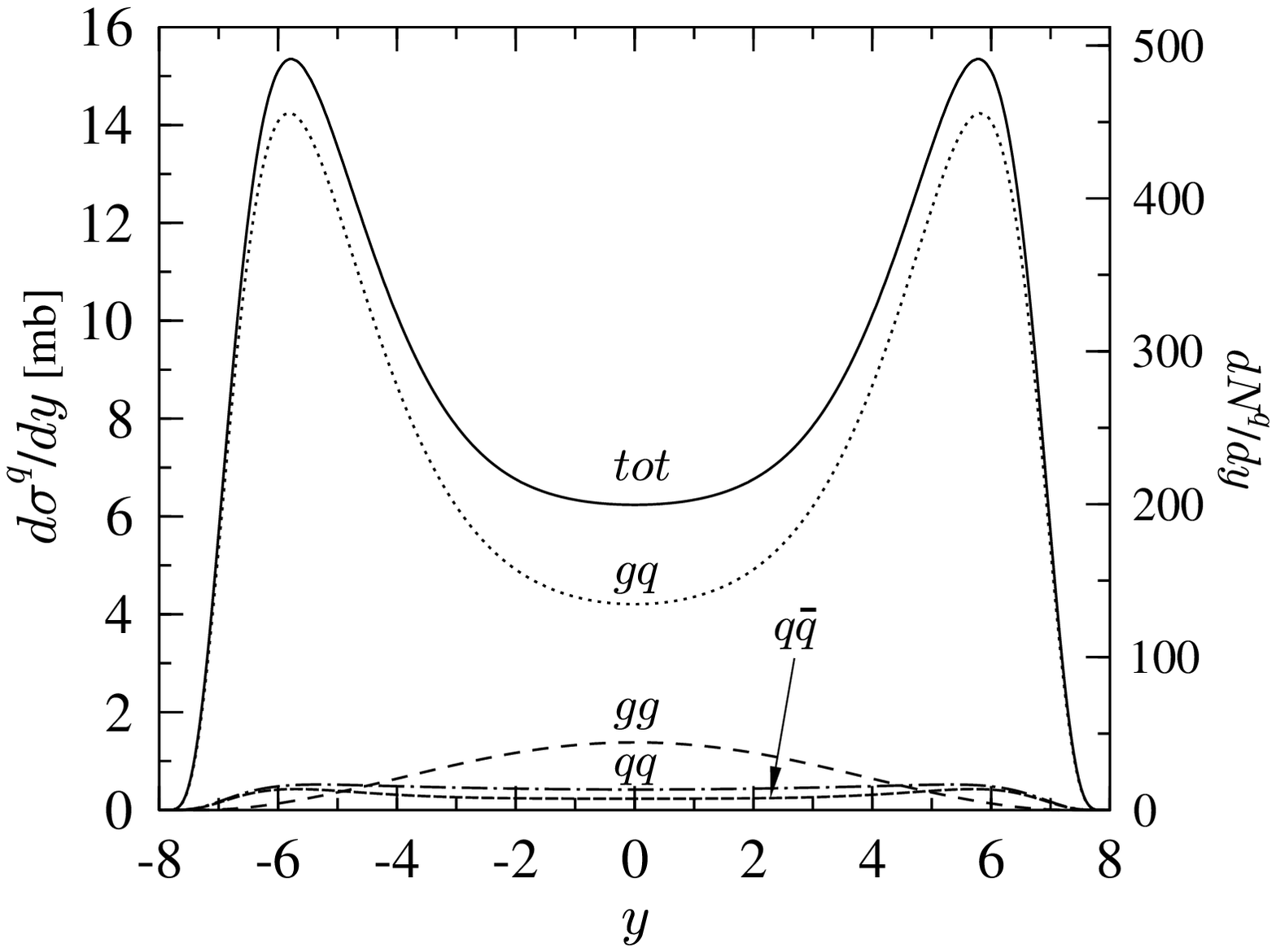}
\hspace{-1cm}
\epsfxsize=10cm\epsfbox{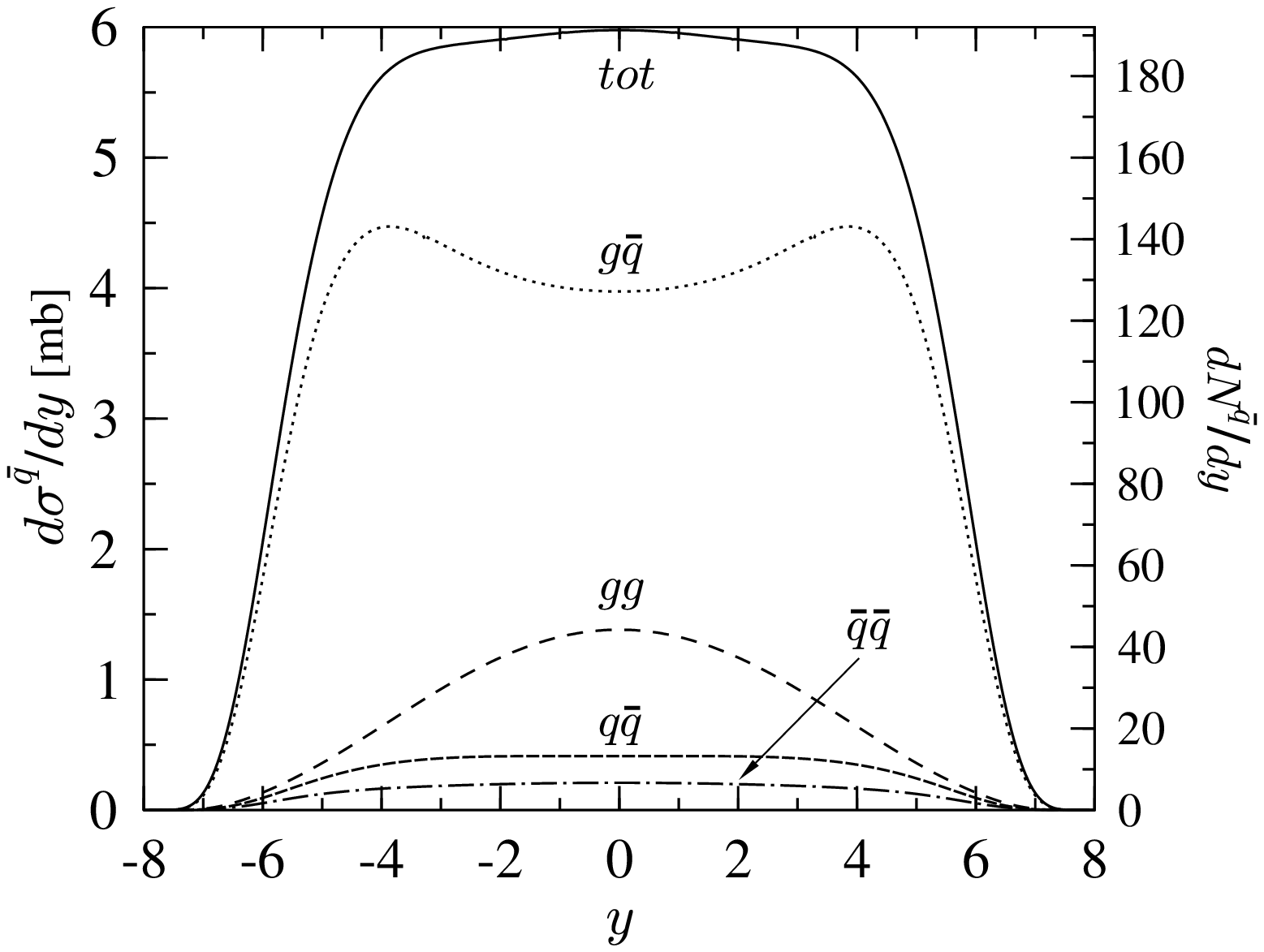}}
\vspace*{-5cm}
\caption[a]{The rapidity distributions of
gluons, quarks and antiquarks ($p_{\rm T}>p_0=2$ GeV) produced in
pp collisions at $\sqrt{s}$ = 5500 GeV
together with the subprocess contributions. 
The axis to the right (to be applied only for $|y|\lsim4$)
gives the corresponding number of partons produced 
in an average central Pb+Pb collision.}
\la{sub5500}
\end{figure}

\section{The results}
The following results are based on a numerical evaluation
of Eqs. (\ref{rate}) and (\ref{sET}). For definiteness, 
we give the results for all $y$,
in spite of the fact that very large $|y|$ phenomena cannot
be experimentally observed in the collider mode and in spite
of the fact that they are theoretically less well founded than
the results near $y=0$ -- they involve structure functions at
extremely small $x\approx p_0^2/s$, where they have not yet
been measured, for nuclei, in particular.

Firstly, to give an idea of the subprocesses contributing 
to $g$, $q$ and $\bar q$ production, 
Fig. 1 gives the $y$ distributions from the various
subprocesses at $\sqrt{s}=5500$ GeV. The scale at left refers to
pp collisions, that on the right, obtained by using 
Eqs.(\ref{number})-(\ref{overlap}), gives the corresponding 
number of partons produced in an average central Pb+Pb collision. 
This scaling from pp to Pb+Pb is not valid for large $|y|$,
where it may lead to a violation of energy or baryon number
conservation. For kinematical reasons very small values of
$x$ are involved (see Fig. 4 below) and shadowing becomes important.
In Fig. 2 for \roots = 200 GeV one is far from violating 
energy or baryon number
conservation in Pb+Pb for $|y|\lsim4$ and one can thus apply
the right hand scale at least up to this $y$; for pp there is
no problem at any $y$. 

\begin{figure}[tb]
\vspace*{2cm}
\centerline{\hspace{-3.3mm}
\epsfxsize=10cm\epsfbox{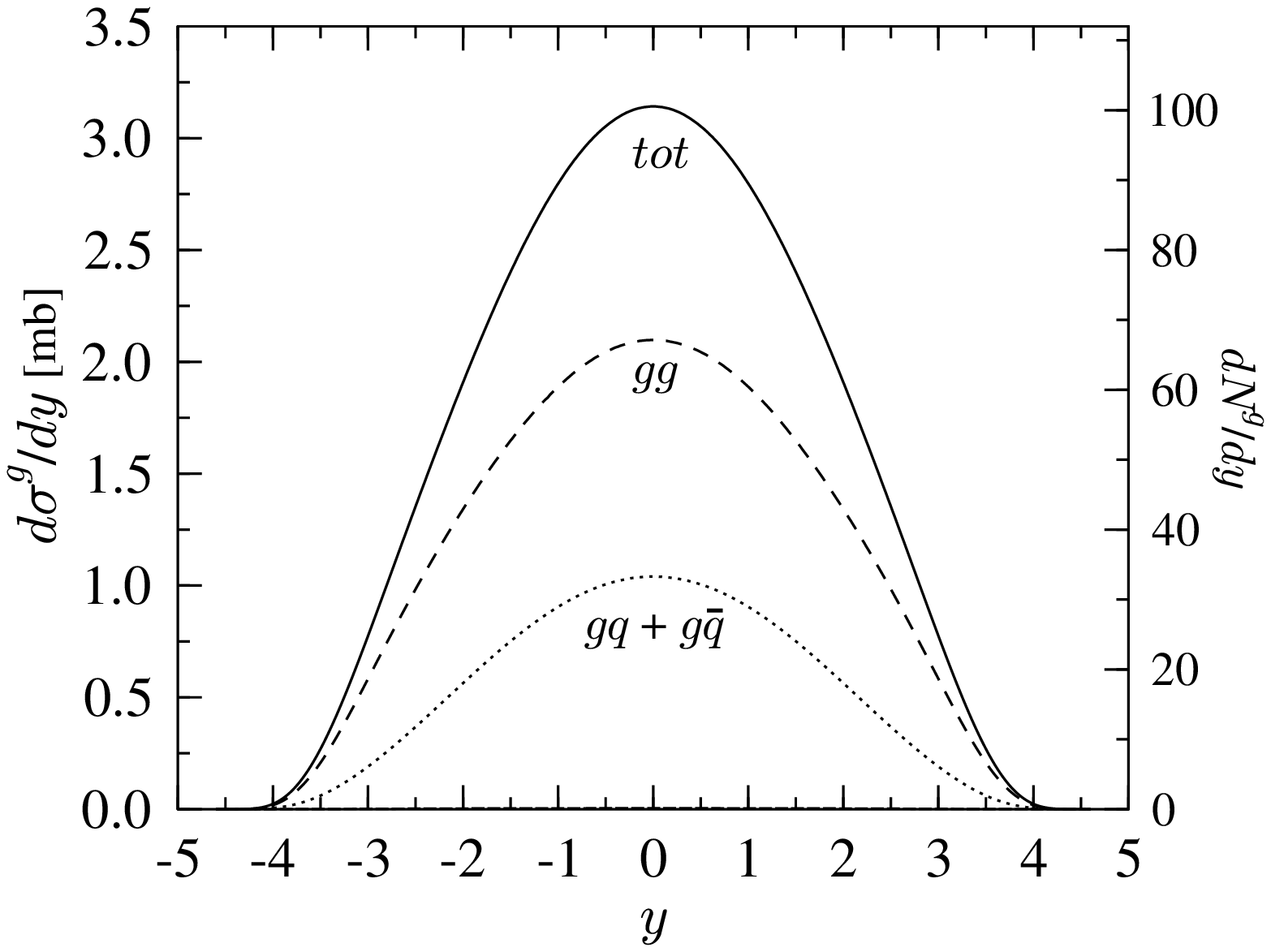}
}
\vspace{-2.4cm}
\centerline{\hspace{-3.3mm}
\epsfxsize=10cm\epsfbox{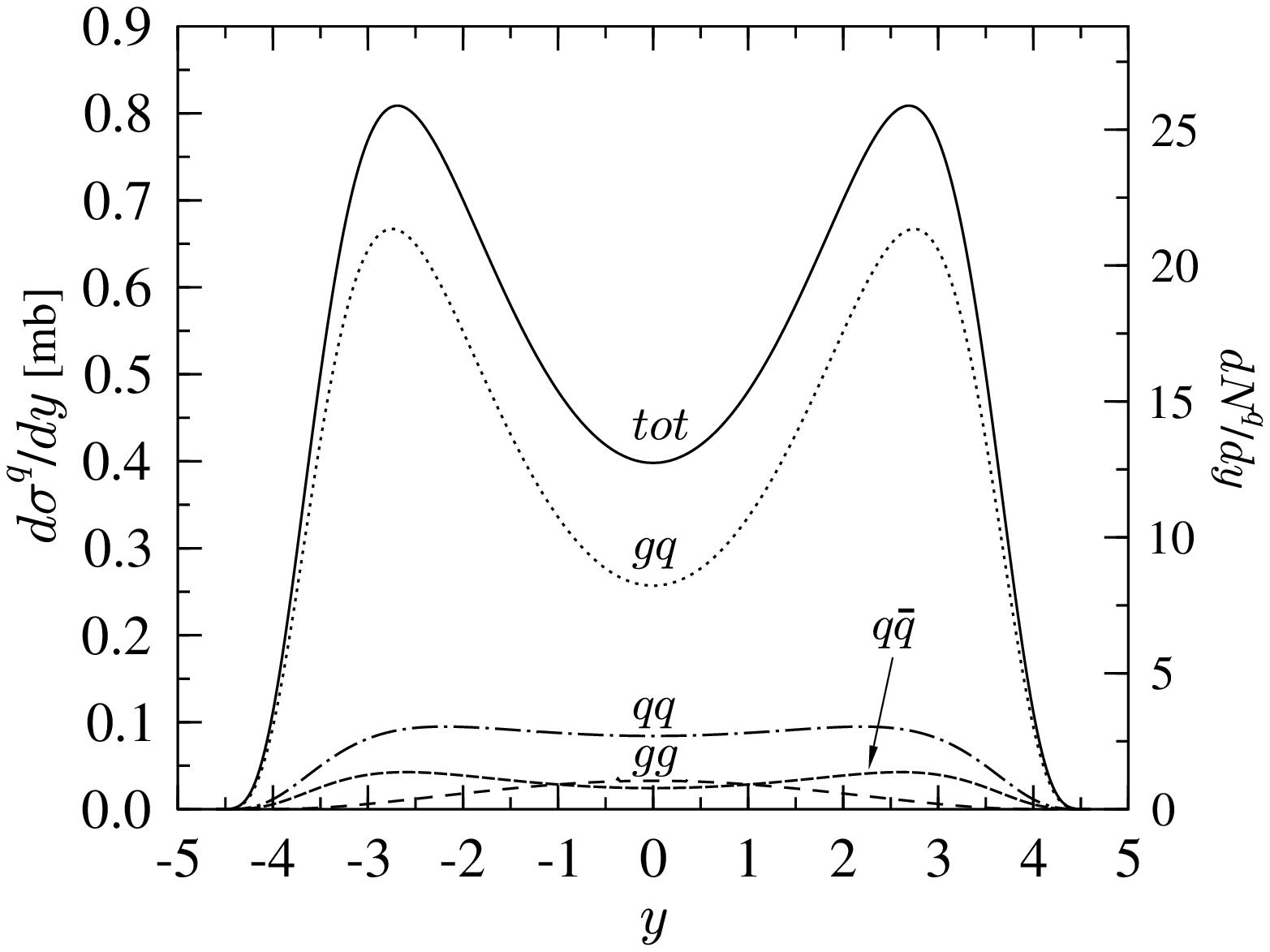}
\hspace{-1cm}
\epsfxsize=10cm\epsfbox{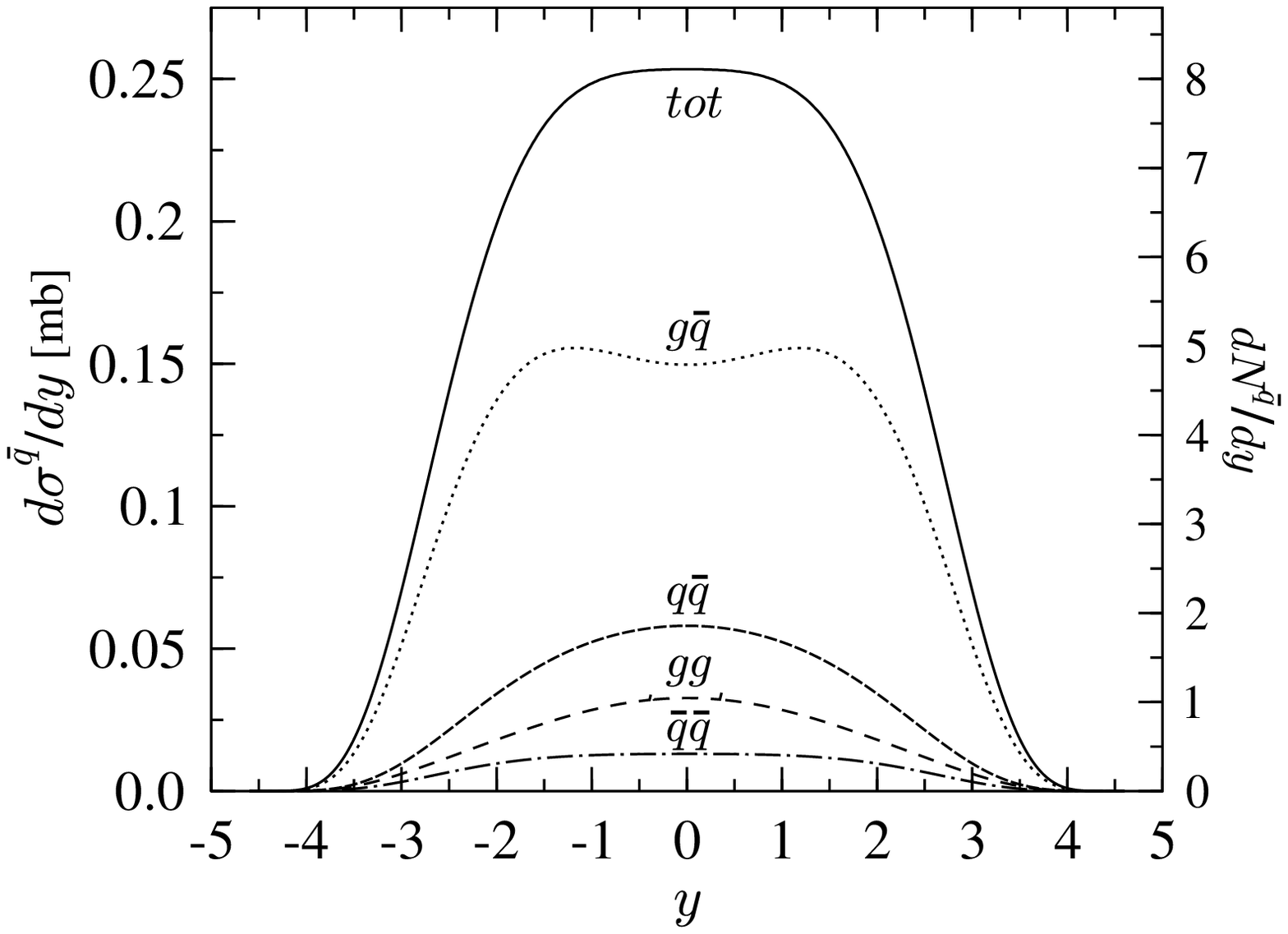}}
\vspace*{-5cm}
\caption[a]{As Fig. 1 but for $\sqrt{s}$ = 200 GeV}
\la{sub200}
\end{figure}

One observes that for gluons the process $gg\to gg$ is dominant
and increasingly so with increasing $\sqrt{s}$. For quarks the 
dominant process is $gq\to gq$ and the forward-backward peaking 
is due to valence quarks in this subprocess. For antiquarks the 
dominant process is $g\bar q\to g\bar q$, without any valence
quark contribution.

The $y$ distributions are summarised in Fig. 3, which also shows
the important net quark number distribution. Some numerical values
are given in Tables 1 and 2.

\begin{figure}[tb]
\vspace*{2cm}
\centerline{\hspace{-3.3mm}
\epsfxsize=10cm\epsfbox{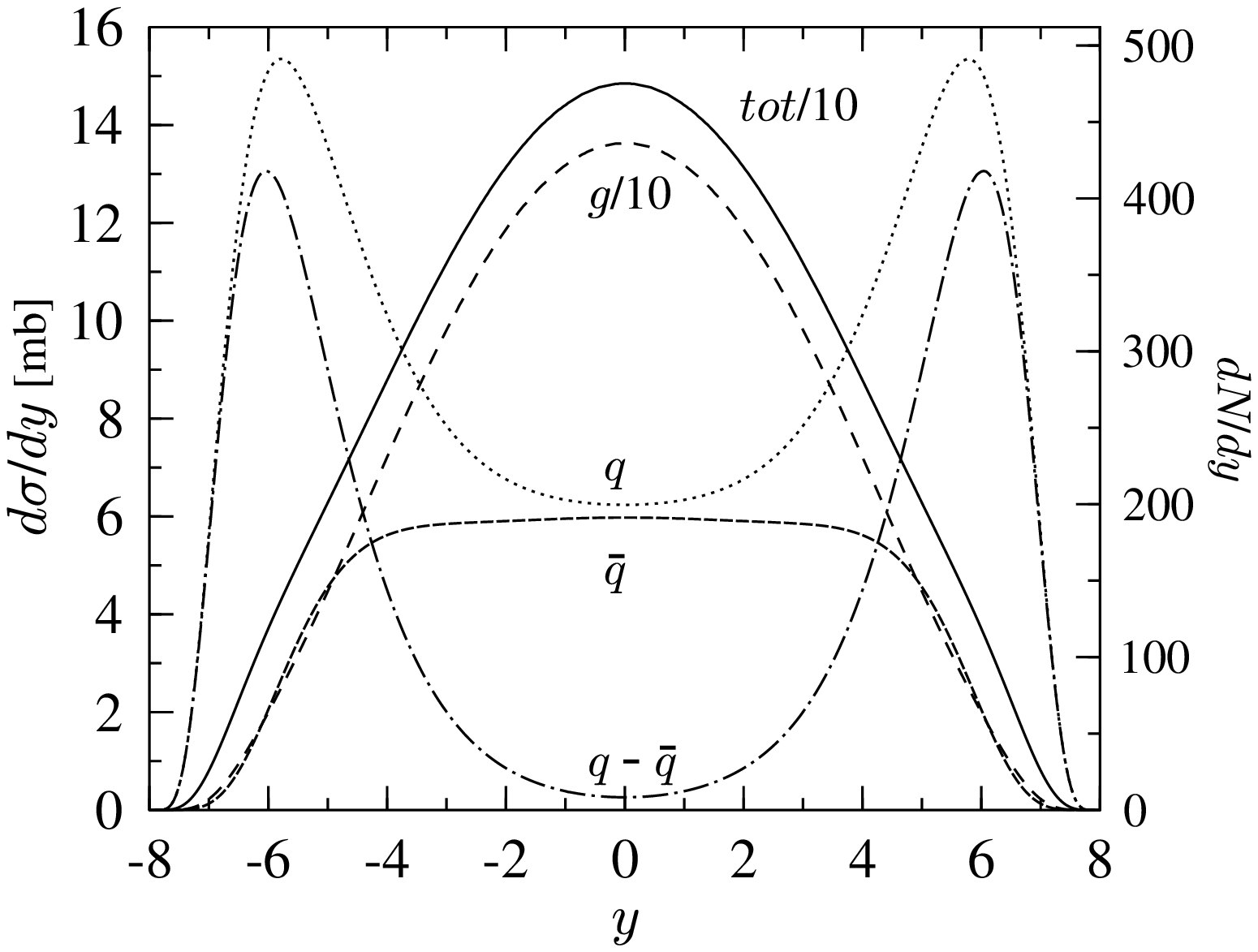}
\hspace{-1cm}
\epsfxsize=10cm\epsfbox{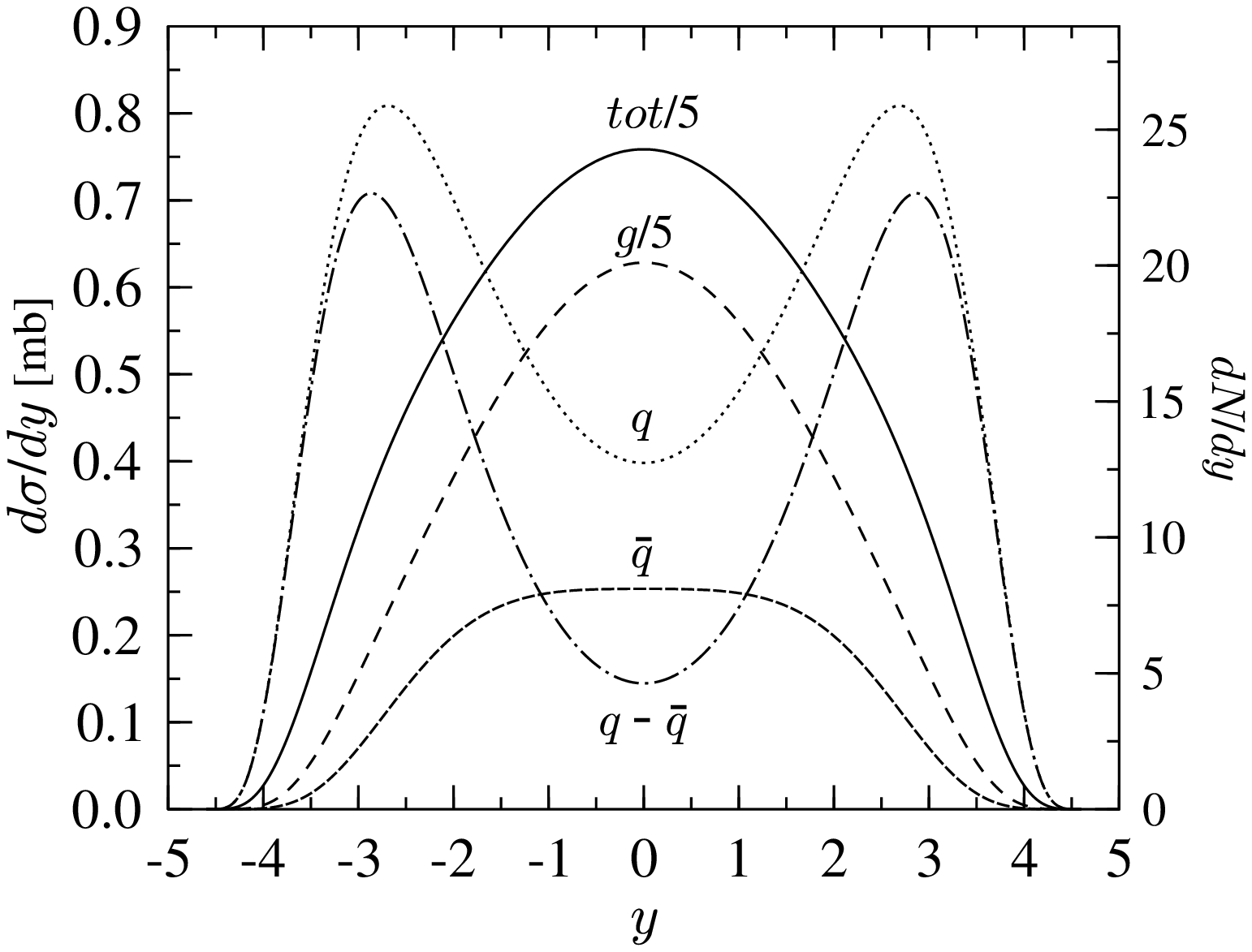}}
\vspace*{-5cm}
\caption[a]{Distributions of various quanta for $\sqrt{s}$
= 5500, 200 GeV. Note that the total hard and gluon cross sections 
have been scaled down by a factor 10 and 5 for $\sqrt s=5500$ GeV 
and 200 GeV, respectively.}
\la{all}
\end{figure}

\vspace{2cm}
\subsection{Initial state}
On the basis of the previous results one can immediately discuss
the appearance of the system at the initial time 
$\tau_i$ = 0.1 fm. For concreteness we give fixed numbers 
neglecting possible errors -- and keep even too many decimals.
Due to the expected dominance of semihard processes in the TeV 
energy  range, let us mainly focus on the results for LHC. 
In the figures and in the tables, we give the numbers for 
$\sqrt s=200$ GeV for comparison, so that our discussion 
may be reproduced for RHIC as well.
Thus, unless otherwise stated, the numbers apply to
an average central Pb+Pb collision at \roots = 5500 GeV, $p_0=2$
GeV and the rapidity range $-0.5<y<0.5$. 
The results in Fig. 1 and Tables 1 and 2 firstly imply that  there are 
\be
4350\,\,{\rm gluons}\quad+\quad 200\,\,{\rm quarks}\quad
+\quad 190\,\,{\rm antiquarks},
\ee
which carry a transverse energy of
\be
12960\,\gev\,\,{\rm for\,\,gluons}\quad+
\quad 620\,\gev\,\,{\rm for\,\,quarks}\quad
+\quad 590\,\gev\,\,{\rm for\,\,antiquarks}.
\ee
In a space-time picture these are formed at a time $1/p_0$ =
0.1 fm after the collision and thus, using eqs.(\ref{radii}),
the corresponding number and energy densities are 
\be
n_g={325\over\fm^3},\quad n_q={14.9\over\fm^3},
\quad n_{\bar q}={14.2\over\fm^3},
\la{numberdensities}
\ee
\be
\epsilon_g=967{\gev\over\fm^3},\quad \epsilon_q=46.3{\gev\over\fm^3},
\quad \epsilon_{\bar q}=43.9{\gev\over\fm^3}.
\la{energydensities}
\ee

\begin{table}
\center
\begin{tabular}{|c|c|c|c|c|c|}
\hline
$\sqrt s/A$ & range of $y$ & total  & $g$   & $q$    & $\bar q$   \\
\hline
5500        & $|y|<0.5$    & 74.06  & 67.95 & 3.123  & 2.988      \\
            & all $y$      & 657.5  & 555.8 & 68.34  & 33.06      \\
\hline 
\hline 
200         &$|y|<0.5$     & 1.885  & 1.556 & 0.2025 & 0.1266     \\      
            & all $y$      & 10.00  & 7.077 & 2.278  & 0.6454     \\

\hline
\end{tabular}
\caption[1]{Values in mb of the perturbative parton cross sections 
 $\sigma_\rmi{hard}^f(\sqrt s,p_0)$ and 
$\sigma_\rmi{hard}^f(\sqrt s,p_0,|y|\le0.5)$
in pp collisions at \roots = 5500 GeV, $p_0=2$ GeV. The number of
partons of type $f=g,q,\bar q$ in an average Pb+Pb collision is
$\bar N^f_{\rm PbPb}=2\sigma^f_\rmi{hard}T_\rmi{PbPb}({\bf 0})
\approx 64\sigma_\rmi{hard}^f$.}
\la{sig}
\end{table}

\begin{table}
\center
\begin{tabular}{|c|c|c|c|c|c|}
\hline
$\sqrt s/A$ & range of $y$  & total & $g$   & $q$    & $\bar q$  \\
\hline
5500       & $|y|<0.5$      & 443   & 405   & 19.4   & 18.4      \\
           & all $y$        & 3830  & 3227  & 407    & 196       \\
\hline 
\hline 
200        & $|y|<0.5$     & 10.02  & 8.211 & 1.127  & 0.6834    \\
           & all $y$       & 51.99  & 36.53 & 12.09  & 3.370     \\

\hline
\end{tabular}
\caption[1]{As Table 1 but for values of the first $E_{\rm T}$-moments 
$\sigma_\rmi{hard}\langle E^f_{\rm T}\rangle$ in mbGeV
of the parton distribution. The transverse energy of
partons of type $f$ in an average central Pb+Pb collision is
$\bar E_{\rm T}^f=\sigma_\rmi{hard}\langle E_{\rm T}^f\rangle T_\rmi{PbPb}({\bf 0})
\approx32\sigma_\rmi{hard}\langle E_{\rm T}^f\rangle$.}
\la{sigET}
\end{table}

Note that here we consider an idealized (simplified) picture of 
a very high energy heavy ion collision, where the
transit time of the colliding nuclei is reduced to zero. For the 
TeV energy range, this is a good approximation, since then 
$2R_{\rm Pb}/\gamma\ll 1/p_0$. However, semihard and softer 
gluons extend longitudinally  much further than the strongly 
contracted valence components. For the semihard partons, 
the typical uncertainty in the  longitudinal direction is  
$\Delta z \sim 2/(x_{1(2)}\sqrt s)\sim 1/p_{\rm 0}$. This width 
in the production mechanism \cite {ew94} is also neglected in the 
idealized picture. Then, in the central rapidity unit, one may estimate 
$\langle E\rangle \approx 2\langle E_{\rm T}\rangle\sinh 0.5 
= 1.04\langle E_{\rm T}\rangle$  for the initial parton production, so that 
converting the transverse energy directly into energy density 
is a good approximation.

Let us briefly recapitulate the discussion of the gluonic
subsystem \cite{ekr}, which has two notable properties.
Firstly, the $E_{\rm T}$ per gluon is
\be
{\bar E^g_{\rm T}\over \bar N_{\rm PbPb}^g}
= \frac{\epsilon_g^{\rm pQCD}} {n_g^{\rm pQCD}}
= 2.98\,{\rm GeV},
\ee
which naturally is somewhat larger than $p_0$. 
For an ideal gas of massless gluons in a complete thermal 
equilibrium,  the temperature can be determined from
\be
\epsilon_g^{\rm ideal}=3\pi^2/90\cdot 16T_{\rm eq}^4.
\ee
For an ideal gas with $\epsilon_g^{\rm ideal}=\epsilon_g^{\rm pQCD}$
we find $T_{\rm eq}=1.10$ GeV.  It is interesting,  and surprising, 
to see that
\be
\frac{\epsilon_g^{\rm pQCD}}{n_g^{\rm pQCD}}  \approx
\frac{\epsilon_g^{\rm ideal}}{n_g^{\rm ideal}} = 2.7 T_{\rm eq}.
\ee
As far as energy per particle goes, the initial
gluon subsystem thus is ``thermalised'' from the beginning. 
Note especially that we have not considered isotropisation of 
the system at all. In our idealized picture, 
 $\langle p_z^2\rangle \ll \langle p_{\rm T}^2\rangle/2$
initially in the central rapidity unit and collisions are needed 
to make the system isotropic. What our conclusion does indicate, 
however, is that since the appropriate energy/gluon is already 
there, no additional particle production is necessary, increasing 
thus possibilities for a rapid thermalisation. At this point it 
is perhaps worth mentioning that both the time of isotropisation 
and the conclusion of thermalisation depends on modelling of the 
initial collisions, especially on how the longitudinal width of 
initial parton production is accounted for. How rapidly 
isotropisation then proceeds depends on the expansion of the system 
and on the rate of secondary collisions. 
Note that if there exists a finite width in the longitudinal 
direction, one may estimate the isotropisation time (and the initial 
conditions) also in a free streaming picture \cite{bdmtw,ew94}.

Secondly, with the present $p_0=2$ GeV the gluonic subsystem  satisfies
precisely the transverse saturation criterion (\ref{saturation})
for $\sqrt s=5500$ GeV. In general, for pp collisions, 
$p_0$ is a phenomenological
parameter separating the  computable perturbative physics from the 
more model dependent soft physics. The value of $p_0$ depends 
on what is assumed for the soft physics, and it is restricted by
the measured total and inelastic cross sections \cite{wang91} and transverse 
momentum spectra of charged particles in p$\bar{\rm p}$ and pp collisions
\cite{sz}. As we have shown above, sufficiently many semihard partons 
are produced  in a heavy ion collision at very high energy, so that 
$p_0$ gets  a better defined {\it dynamical} significance, indicating 
that the soft component becomes less relevant. 
From Table 1 we can see that there is no
transverse saturation at $p_0=2$ GeV for $\sqrt s=200$ GeV but the 
saturation occurs for smaller $p_0$, involving $p_{\rm T}$'s 
in the non-perturbative regime. In fact, saturation
is reached at $p_0=1.6$ GeV for \roots=1.8 TeV and at
$p_0=1.0$ GeV for \roots=200 GeV. 
Therefore, at the RHIC energies, one should 
not neglect the soft component.
A phenomenological way to obtain the saturation
(dynamical screening) in a self-consistent manner in the initial 
parton production  was suggested in Ref. \cite{emw}.

Consider then the (anti)quarks, the density of which is below the
thermal density by a factor of about 30. From 
Eq.(\ref{numberdensities}) one sees that the initial net
baryon number density at the time $1/p_0=0.1$ fm is
\be
n_{B-\bar B}\equiv\frac{1}{3}(n_q-n_{\bar q}) = {0.21\over\fm^3}.
\ee
Thus even at these ultrarelativistic energies the initial net
baryon number density is more than the usual nuclear matter
density, even if the initially baryon-rich fragmentation regions 
are already far apart in the forward and backward directions.
Similarly, one may also compute the initial
asymmetry:
\ba
{B-\bar B \over B+\bar B}&=& 18\%....29\%,\qquad\sqrt{s}
=200\,\gev,\nonumber\\
&=& 4.7\%...6.3\%,\qquad\sqrt{s}=1800\,\gev,\la{asymmetry}\\
&=& 2.1\%...2.6\%,\qquad\sqrt{s}=5500\,\gev,\nonumber
\ea
where the range of values shown corresponds to varying $p_0$
from 1 to 3 GeV and where the result also is given for lower
energies for comparison. Note that the initial asymmetry depends
only mildly on $p_0$. The difference between $\sqrt s=$ 5500 GeV 
and 200 GeV is clearly due to the increase in the quark and antiquark 
rates with increasing energy, the difference $B-\bar B$ 
varying less rapidly.  Although the mechanism considered here is 
perturbative, the initial asymmetry for the LHC energies
is of the same order of magnitude as what is obtained in Ref. 
\cite{kharzeev} in a purely non-perturbative gluon field model.

Finally, since for a thermal boson gas $s=3.60n$, the total
initial entropy (using the number of effectively thermal gluons)
is $S=15900$ and, since $B-\bar B=2.9$ initially, the initial
net baryon-to-entropy ratio at $\tau_i=0.1$ fm is
\be
{B-\bar B\over S}\approx1.8\cdot10^{-4}\sim{1\over5000}. 
\label{boversinitial}
\ee
It is also interesting to compare this number with the one for
the early Universe, where the inverse of the specific entropy 
is $\sim 10^{-9}$. Thus we are still relatively far from the 
extreme conditions in the early Universe.

\subsection{Further evolution}
Further evolution of the system will go through possible 
thermalisation (quarks also may be thermalised), 
plasma expansion, phase transition, hadron gas expansion
and decoupling stages. It is clear that any conclusions will
at present be model dependent.

The simplest assumption is that of instant thermalisation and
further essentially adiabatic expansion during all the stages, 
in analogy with the early universe. 
If the entropy ($S\approx15900$),
net baryon number ($B-\bar B\approx3$) and the net
baryon number-to-entropy ratio  ($(B-\bar B)/S=1/5000$)
are all conserved, the initial
values are also the final ones. 
For a hadron gas $S/N\gsim4$ depends somewhat on the 
decoupling temperature. For a pion gas $S\approx4N$
and a multiplicity of about 4000 is predicted.

The assumption of instant thermalisation for quarks is, of
course, unrealistic. Instead, one may assume that they thermalise
adiabatically at some later time $\tau_q$. From the
constancy of entropy it then follows that $g_gT^3=(g_g+g_q)
T_\rmi{new}^3$, {\it i.e.},
\be
T_\rmi{new}=T\bigl(1+g_q/g_g\bigr)^{-1/3};
\ee
the increase $g_g\to g_g+g_q$ in the number of thermalised degrees 
of freedom is compensated for by a
decrease in $T$. The total entropy and the multiplicity
prediction is unchanged, but the energy density is decreased
by the same factor as $T$. Note that the adiabatic scenario
of quark thermalisation leads to a lower limit of the final pion 
multiplicity. Non-equilibrium models, where entropy, and therefore 
also the pion multiplicity, increase due to the chemical 
equilibration of quarks have also been presented
\cite{bdmtw}.

Assuming that each produced particle carries on the average
a transverse energy of $m_T\approx 0.5$ GeV, the final 
transverse energy in the thermal scenario is about 2000 GeV.
What happened to the large initial energy of 13000 GeV in 
Eq.(\ref{energydensities}) in this scenario? It has been
transmitted to larger rapidities by the work done against
longitudinal expansion. In other words, the total entropy
and number of particles in a unit of rapidity  remain constant,
but since the energy/particle $\sim T\sim 1/\tau^{1/3}$
decreases, the total energy in a unit of rapidity decreases. 
The total reduction factor is the ratio between initial
and final values of energy/particle = $2.7T_i/m_T\approx6$. 

We thus come to a series of scenarios starting from fully adiabatic
expansion proceeding via increased amount of dissipation to
a free-streaming expansion. The predictions of the final
multiplicity and $E_{\rm T}$ starting from the same initial values
in eqs. (\ref{numberdensities},\ref{energydensities}) thus can vary
from $N_\rmi{fin}=4000, E_\rmi{Tfin}=2000$ GeV  to values larger than
these by a factor of up to six.  To narrow this range, reliable 
estimates for mean free paths would be needed.

It may be of some interest to show how increasing dissipation can
be described by the introduction of a shear viscosity
\cite{viscosity,eg}. If one writes $\eta=\eta_0 T^3$  and uses 
$\epsilon=3aT^4$, the solution of the usual longitudinal 
hydrodynamical equations  for $T$ is
\ba
T&=&T_i\biggl({\tau_i\over\tau}\biggr)^{1/3}+
{\eta_0\over3a}{1\over2\tau_i}
\biggl[\biggl({\tau_i\over\tau}\biggr)^{1/3}-
{\tau_i\over\tau}\biggr]\nonumber\\
&\to&\biggl[T_i+{\eta_0\over 6a\tau_i}\biggr]
\biggl({\tau_i\over\tau}\biggr)^{1/3}.
\label{viscousT}
\ea
Dissipation thus increases the temperature one reaches at a
certain proper time and thus it also increases the lifetime
and the multiplicity. All of this is parametrised by
$\eta_0$. By definition, the effect must be small,
$\eta_0\ll 6aT_i\tau_i$. In the present context,
$\eta_0$ could, in principle, be measured by observing deviations
from the adiabatic multiplicity of about $N_\rmi{fin}=4000$.
In practice, however, this leading estimate is not reliable
enough to permit measurements of corrections to it.

\section{Soft component?}
One of the motivations for the choice of taking $p_0=2$ GeV was
that at this value the produced gluons are so numerous that
they are transversally saturated (Eq.(\ref{saturation})) so
that \cite{bm} this semihard component gives a quantitatively
correct estimate of an average event at LHC energies. At lower
energies the saturation is reached for still smaller values
of $p_0$ (at $p_0=1.6$ GeV for \roots=1.8 TeV, at
$p_0=1.0$ GeV for \roots=200 GeV), for which next-to-leading
order corrections are expected to be larger. At SPS energies,
\roots = 20 GeV, the whole approach becomes useless. 
At these energies the bulk of the events is described by a 
nonperturbative soft component based on string or colour field 
formation and decay.

A way to assess the importance of the corrections is to 
assume that the soft component, dominant at \roots = 20 GeV, is
constant \cite{wang91}. The SPS data then  imply \cite{ETdata} 
that this soft component gives an average  contribution of 
$\bar E_{\rm T}^{\rm soft}({\bf 0}) \approx 420 $ GeV 
in the central rapidity unit of central Pb-Pb collisions.
The corresponding perturbative number $\bar E_{\rm T}^{\rm hard}({\bf 0})$ 
is 14 200 GeV at \roots = 5500 GeV ($p_0$ = 2.0 GeV, Table 2).
With these numbers, (assuming that for the SPS the hydrodynamical 
stage of the system is not very long) a simple order-of-magnitude 
estimate for the final baryon-to-entropy ratio can be made. Let us 
decompose the ratio into soft and hard contributions:

\begin{equation}
\frac{B-\bar B}{S}
= p_{\rm hard}\biggl(\frac{B-\bar B}{S}\biggr)_{\rm hard}
+ p_{\rm soft}\biggl(\frac{B-\bar B}{S}\biggr)_{\rm soft},
\end{equation}
where $ p_{\rm hard} = S_{\rm hard}/S$
and $p_{\rm soft} = S_{\rm soft}/S$. An estimate for these
is obtained from the transverse energies above,
\begin{equation}
p_{\rm hard} \approx \frac{\bar E_{\rm T}^{\rm hard}}
{\bar E_{\rm T}^{\rm hard} + \bar E_{\rm T}^{\rm soft}}, \,\,\,\,\,\,
p_{\rm soft} \approx \frac{\bar E_{\rm T}^{\rm soft}}
{\bar E_{\rm T}^{\rm hard} + \bar E_{\rm T}^{\rm soft}}.
\end{equation}

Using then the baryon-to-entropy ratio as in Eq. (\ref{boversinitial})
for the hard component, and 1/50 from Ref. \cite{lthsr} for the 
soft component, we obtain the final baryon-to-entropy ratio
\begin{equation}
\frac{B-\bar B}{S} \sim 8\cdot 10^{-4}
\end{equation}
for nuclear collisions at LHC. 

For RHIC, the corresponding estimate is more unreliable 
due to the following reasons:
Firstly, as discussed above, at \roots = 200 GeV, the saturation of 
minijet production occurs at $p_{\rm T}\sim 1.0$ GeV but as this 
may already be in the region where the minijet cross sections become 
unreliable, let us consider a range from $p_0= 1.5$ GeV to 2.0 GeV.
This gives $\bar E_{\rm T}^{\rm hard}({\bf 0})$ = 708 GeV to 326 GeV.
Secondly, although the gluons do dominate the early parton system 
also at RHIC, the energy/gluon is not thermal, so the assumption 
of instant thermalisation of glue at $\tau\sim 0.1$ fm is 
probably not as good as for the LHC. Following, nevertheless, the same 
line of reasoning as for the LHC gives 
$(B-\bar B)/S\sim 2\cdot10^{-3}$ initially  
at $\tau \sim 0.1$ fm, 
and $(B-\bar B)/S\sim 9\cdot10^{-3}$ 
for the final  baryon-to-entropy ratio for nuclear collisions at RHIC.

We see that the nonperturbative sources may increase the 
baryon-to-entropy 
ratio considerably, by a factor of about 4, from the initial values 
both at LHC and RHIC. 
At the LHC, the increase is dominantly due to the increase in the 
net baryon number and not in the total entropy, while at the RHIC
both the net baryon number and the total entropy are increased by
the soft source contributions. 

\section{Very forward and backward partons}
The previous discussion has been confined to the $y=0$ region,
although the curves in Figs. 1-3 are computed over the entire
rapidity range. They are reliable on the pp level, but the
simple transformation in Eqs. (\ref{number},\ref{overlap}) used to
transform them to the A+A level assumes independent subcollisions
and thus does not take into account
energy and baryon number conservation. It thus fails 
at large $y$, where a few gluonic subcollisions can consume all
the available energy or valence quark+gluon subcollisions can
consume all the baryon number. For the approach to be valid
one has to be far from these limits; there is no problem at
RHIC, but at LHC one should not go beyond $|y|\approx4$.

Another problem is that kinematically a very forward parton demands
that the other one also be at large $y$. This, on the other hand,
implies that the $x$ value of one of the partons is extremely
small and in a region in which the structure functions
are unmeasured. This effect is quantitatively discussed in 
Fig. 4.

\section{Conclusions}
We have in this paper computed the initial number of gluons, quarks
and antiquarks with $p_{\rm T}>p_0$ = 2 GeV (value determined by
transverse saturation) 
produced in Pb+Pb collisions at \roots = 5500 GeV near $y=0$ and
studied the further evolution of this system especially in view of
the behavior of the net baryon number. The main conclusions are that
the produced gluon system is transversally saturated and the parameter 
$p_0$ gets a dynamical significance in LHC nuclear collisions. 
The perturbatively produced gluon system has also energy/gluon
as in a thermal ideal gas, so only isotropisation is needed to 
make the system  thermalised. This clearly enhances possibilities 
of a very rapid thermalisation. The initial net baryon number density at 
$\tau_i = 0.1$ fm  may be of the order of nuclear matter density, 
or even larger, but this is rapidly diluted. Initially at $\tau_i=0.1$ fm
the baryon-to-entropy ratio is  $(B-\bar B)/S\sim 1/5000$, and
the final baryon-to-entropy ratio may be larger by up to a factor 4.
We conclude that the baryon-to-entropy ratio is small, though still 
orders  of magnitude larger than the corresponding quantity in cosmology.

The main sources of uncertainly in the computation are
next-to-leading order corrections, nuclear shadowing of the
initial parton distributions and dissipative effects  and the 
non-perturbative source terms during expansion and hadronisation. 
Next-to-leading order corrections with the acceptance conditions 
applicable here have not been computed but should be. Nuclear 
gluon shadowing can only be modelled and ultimately has to be 
measured. This also holds for dissipative effects.  Although the 
detailed numbers are uncertain, we believe that this general 
framework will be used to analyse the data when it finally comes.

\begin{figure}[tb]
\vspace*{2cm}
\centerline{\hspace{-3.3mm}
\epsfxsize=10cm\epsfbox{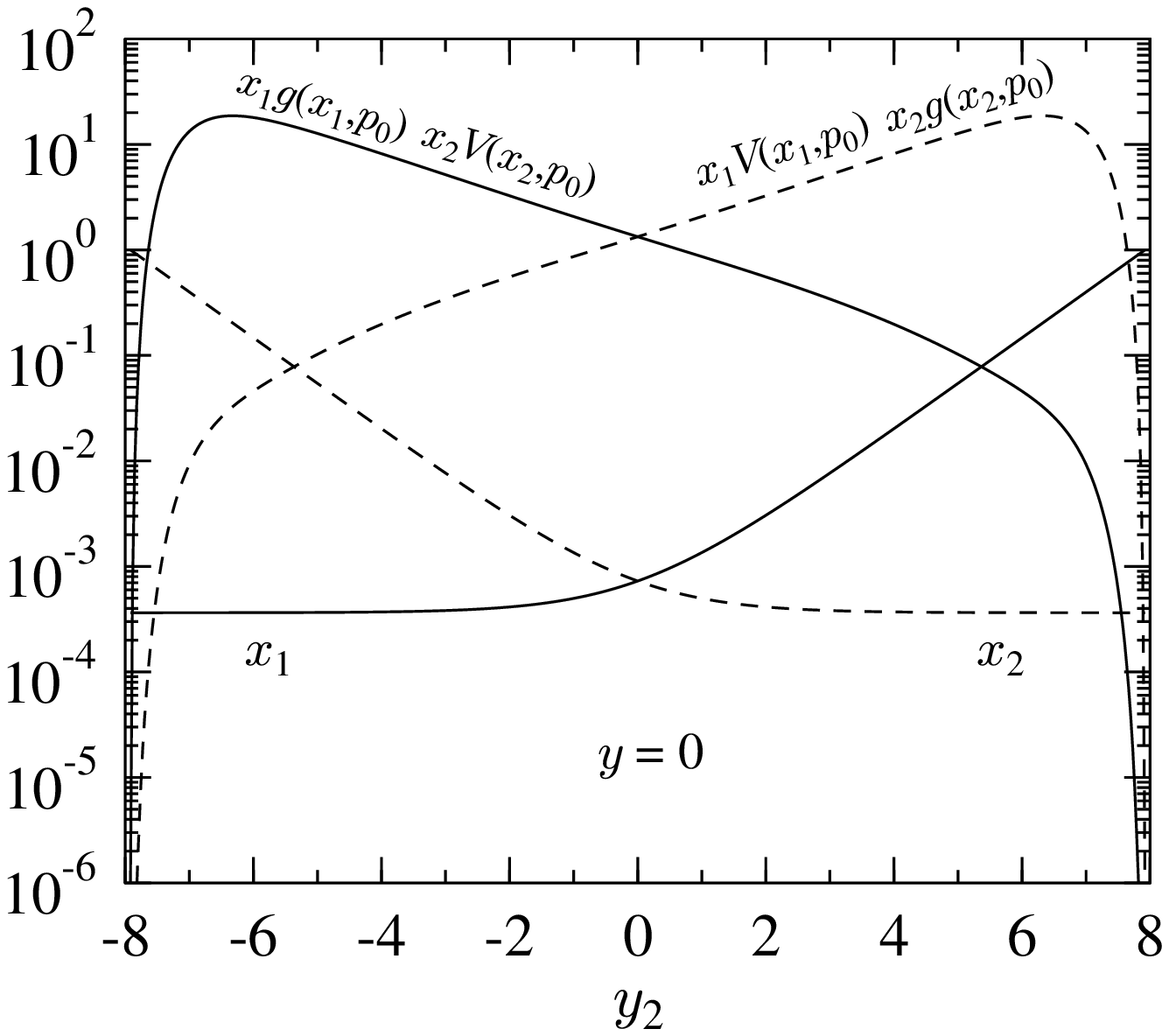}
\hspace{-1cm}
\epsfxsize=10cm\epsfbox{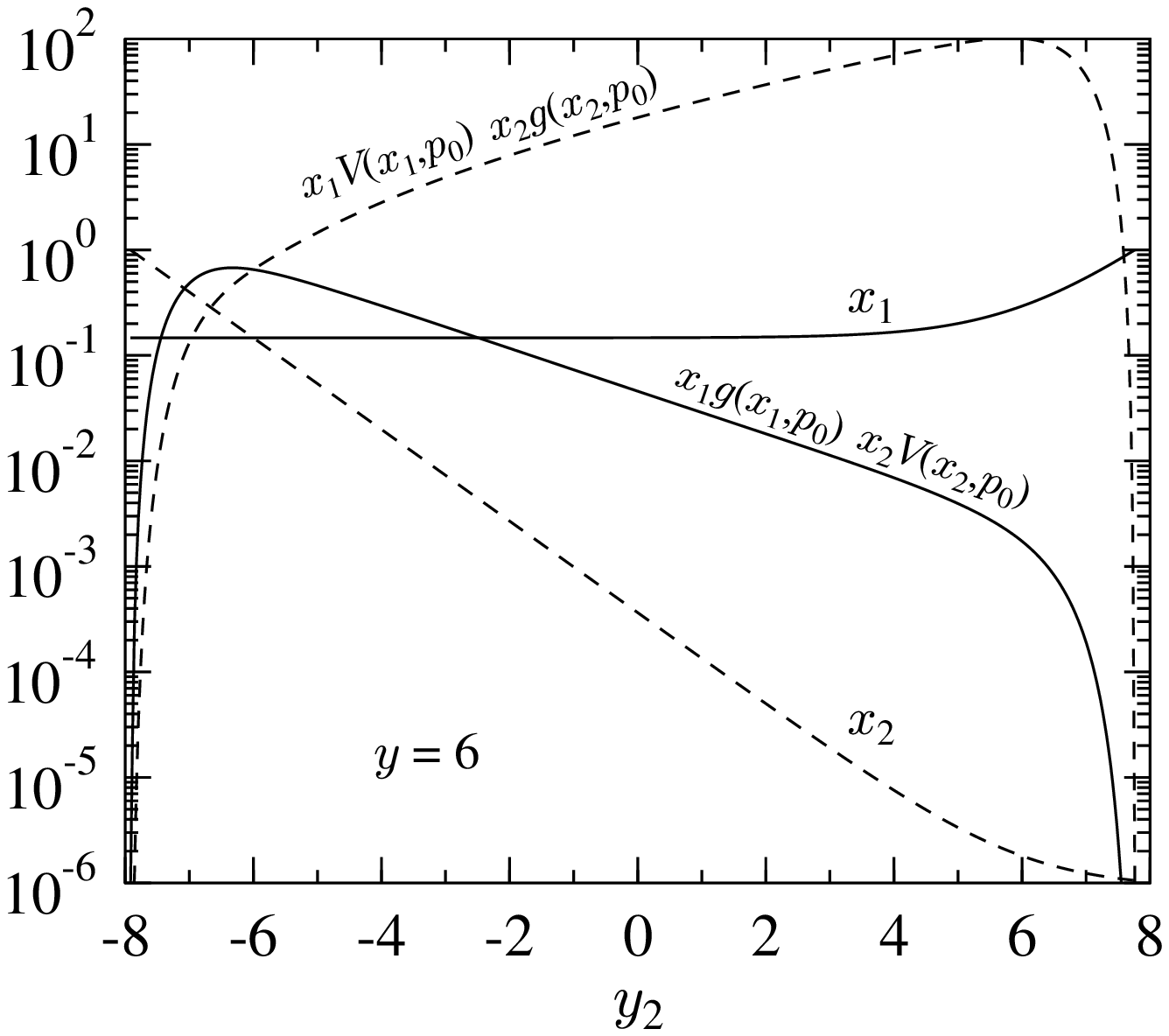}}
\vspace*{-5cm}
\caption[a]{The forward-backward peaking of the quark distribution
in Fig. 1 arises from valence-gluon subcollisions. This figure shows
the parton luminosities $x_1g(x_1,p_0)x_2V(x_2,p_0)$ and
$x_1V(x_1,p_0)x_2g(x_2,p_0)$ as a function of the rapidity $y_2$ of 
the gluon, which is integrated over in Eq. (\ref{rate}).
The $x_1$ and $x_2$ values of the initial valence quark or gluon
are also shown as a function of the rapidity $y_2$. 
For very forward quark rapidities ($y=6$) 
the dominant configurations are such in which also the final gluon
is at $y_2\approx6$, but then also the $x_2$ is extremely
small ($\sim p_0^2/s$) and in a region where the structure
functions are unknown.
}
\la{lumi }
\end{figure}

\bigskip
{\bf Acknowledgements} We thank D. Kharzeev for reviving our interest 
in the subject and M. Gyulassy, U. Heinz, L. McLerran, V. Ruuskanen,
G. Schuler, and H. St{\"o}cker for useful discussions.

\end{document}